\documentclass[conference]{IEEEtran}
\IEEEoverridecommandlockouts
\usepackage{cite}
\usepackage{amsmath,amsthm,amssymb,amsfonts,bm}

\newtheorem{lemma}{Lemma}

\newtheorem{prop}{Proposition}

\newtheorem{assmp}{Assumption}
\newtheorem{defn}{Definition}
\usepackage{algorithmic}
\usepackage{graphicx}  
\usepackage{epstopdf}
\usepackage{textcomp}
\usepackage{xcolor}
\usepackage{url}
\usepackage{enumitem}
\usepackage{url}
\usepackage{float}
\setlist[itemize]{leftmargin=*}
\setlist[enumerate]{leftmargin=*}
\usepackage{multirow}
\usepackage{bbm}

\usepackage{subfigure}
\usepackage[linesnumbered, ruled]{algorithm2e}

\def\BibTeX{{\rm B\kern-.05em{\sc i\kern-.025em b}\kern-.08em
    T\kern-.1667em\lower.7ex\hbox{E}\kern-.125emX}}
\title{
Regulating Clients’ Noise Adding in Federated Learning without Verification
}

\author{\IEEEauthorblockN{Shu Hong}
\IEEEauthorblockA{
\textit{Singapore University of Technology and Design, Singapore}\\
shu\_hong@mymail.sutd.edu.sg}
\and
\IEEEauthorblockN{ Lingjie Duan
}
\IEEEauthorblockA{
\textit{Singapore University of Technology and Design, Singapore}\\
lingjie\_duan@sutd.edu.sg}
}

\begin{document}

\setlength{\abovedisplayskip}{2pt}
\setlength{\belowdisplayskip}{2pt}

\maketitle

\begin{abstract}
In federated learning (FL), clients cooperatively train a global model without revealing their raw data but gradients or parameters, while the local information can still be disclosed from local outputs transmitted to the parameter server. With such privacy concerns, a client may overly add artificial noise to his local updates to compromise the global model training, and we prove the selfish noise adding leads to an infinite price of anarchy (PoA). 
This paper proposes a novel pricing mechanism to regulate privacy-sensitive clients without verifying their parameter updates, unlike existing privacy mechanisms that assume the server’s full knowledge of added noise. Without knowing the ground truth, our mechanism reaches the social optimum to best balance the global training error and privacy loss, according to the difference between a client's updated parameter and all clients’ average parameter. We also improve the FL convergence bound by refining the aggregation rule at the server to account for different clients’ noise variances. Moreover, we extend our pricing scheme to fit incomplete information of clients’ privacy sensitivities, ensuring their truthful type reporting and the system’s ex-ante budget balance. Simulations show that our pricing scheme greatly improves the system performance especially when clients have diverse privacy sensitivities.

\end{abstract}








\section{Introduction}

The proliferation of Internet of Things (IoT) technologies has led to the creation of enormous amounts of user data. In an effort to safeguard data privacy, decentralized machine learning approaches such as Federated Learning (FL) \cite{kairouz2021advances} have been proposed without sharing individuals' raw data. During the training phase, each client periodically downloads the global model from a parameter server and updates their local model with their own data. By only transferring gradient parameters instead of raw data, data privacy can be preserved.

Sharing model updates or gradients, however, can also compromise clients' private information, as analyzing the differences in training parameters uploaded by clients can reveal it. For example, model inverse attack \cite{geiping2020inverting,zhao2020idlg} allows training data reconstruction by matching model gradients and optimizing randomly initialized inputs. These attacks for data reconstruction or membership inference \cite{shokri2017membership} are significant threats to FL security and privacy. To prevent privacy leakage, each client can add artificial noise \cite{abadi_deep_2016,wei_federated_2020,yang2022accuracy} to the transmitted parameters. However, this approach can lead to significant training accuracy loss while making it harder for attackers to inverse the original local data.


Despite of the privacy concerns in FL, clients might also take advantage of the returned global model for benefits. For example, in the ``cross-silo'' scenario of FL \cite{kairouz2021advances} (e.g., in medical and financial institutions), an AI model is trained on 
clients with large amounts of data and clients also care about the convergence performance.
Given both privacy concerns and model training benefits, Noble et al. \cite{noble2021differentially} and Sun et al. \cite{sun2021pain} studied the tradeoff between utility and privacy for FL.
It is also critical to study the clients' selfish noise-adding strategy based on each individual's preference of learning performance and privacy requirement. 

Our analysis of the worst-case ratio of social costs under clients' selfish behavior and the social optimum prompts us to study incentive mechanisms to regulate clients' noise adding and approach the social optimum. Existing works on noise-adding privacy protection mechanisms \cite{abadi_deep_2016,wei_federated_2020,yang2022accuracy}) assume that the server can verify and know all clients' contributions, which is challenging to achieve in practice without access to clients' local data. In contrast, our paper incentivizes honest contributions from clients with performance-based pricing schemes that penalize individuals based on the difference between their local parameters and the average parameter benchmark of all clients. Although the average benchmark may not be the ground truth, we prove its effectiveness in preventing clients from adding excessive local noise.

Specifically, considering clients' varying privacy sensitivities, we analyse their equilibrium strategies under a properly improved weighted aggregation rule, and optimally design the price charged to each client for achieving the social optimum.
Additionally, we extend our pricing scheme to account for incomplete information about clients' privacy sensitivity, where some clients may not truthfully report their types. Our refined pricing mechanism includes a reward function, a penalty function, and a compensation payment to ensure truthful reporting, social efficiency, and FL system's ex-ante budget balance. Simulation results demonstrate significant improvement in equilibrium performance, particularly when clients have diverse privacy sensitivities.

The rest of this paper is structured as follows. 
In Section II we introduce the system model and problem formulation. 
In Sections III and IV we present the equilibrium analysis and pricing mechanism design under complete and incomplete information of clients' privacy sensitivity types, respectively.
We conclude the paper in Section V.
Due to page limit, we put the detailed proofs in the on-line technical report\cite{TechnicalReport}.

\section{System Model and Problem Formulation}
\label{Sec: noise add}
In this section, we introduce the system model and formulate the privacy-preserving problem with noise adding. 

\subsection{Preliminaries}
First, we present preliminaries about backgrounds on federated learning (FL) and differential privacy (DP).

\textbf{Federated Learning.}
Consider an FL system consisting of one parameter server (PS) and $N$ clients in set $\mathcal{N}=\{1,2,\cdots,N\}$. 
Each client $i \in \mathcal{N}$ holds the local database $\mathcal{D}_i$ and aims to iteratively minimize a local loss function $F_i$ by training the local parameter vector $\mathbf{w}_i$, i.e.,
\begin{eqnarray*}
\mathbf{w}_i=\arg \min_{\mathbf{w}} F_i(\mathbf{w},\mathcal{D}_i).
\end{eqnarray*}
The goal of the server is to learn a global model over data at all $N$ distributed clients.
The problem to minimize the global loss function $F$ is:
\begin{eqnarray}
\label{Equ:global opt prob}
\mathbf{w}^*=\arg \min_{\mathbf{w}} F(\mathbf{w},\mathcal{D}) 
=\arg \min_{\mathbf{w}} \sum_{i \in \mathcal{N}} \rho_i F_i(\mathbf{w},\mathcal{D}_i),
\end{eqnarray}
where $\mathcal{D}=\cup_{i \in \mathcal{N}}\mathcal{D}_i$ includes the datasets from all $N$ clients.
Formally, in each iteration the server aggregates the updated parameters received from all clients as
\begin{eqnarray}
\label{Equ:aggregation general}
{\bar{\mathbf{w}}}(\bm{\rho},\mathbf{w})=\sum_{i\in \mathcal{N}} \rho_i \mathbf{w}_i,
\end{eqnarray}
where $\bar{\mathbf{w}}$ is the parameter vector aggregated at the server and $\rho_i \geq 0$  is the weightage parameter for aggregation.

Each training iteration of such an FL system usually contains the following three steps:
clients' local training and updating, global aggregation at the server, and parameter broadcast to clients.
After a sufficient number of training iterations (say, $T$ rounds) and update exchanges between the server and the distributed clients, the solution to the optimization problem (\ref{Equ:global opt prob}) converges to that of the global model.
Thus, the training accuracy loss can be defined as the expected global error, i.e.,
\begin{equation}
\label{Equ:training accuracy loss}
\mathcal{L}^u=\mathbb{E} \{F({\mathbf{w}}^{(T)})-F({\mathbf{w}}^*)\},
\end{equation}
which depends on the aggregation rule in (\ref{Equ:aggregation general}). 
To quantify the convergence bound for the learning, we here make some typical assumptions about the loss functions. 
\begin{assmp}
The global loss function $F(\cdot)$ is convex and  $L_F$-smooth, i.e., $\left\|\nabla F(\mathbf{x})-\nabla F\left(\mathbf{x}^{\prime}\right)\right\|_2 \leq L_F\left\|\mathbf{x}-\mathbf{x}^{\prime}\right\|_2, \forall \mathbf{x}, \mathbf{x}^{\prime}$. 
\end{assmp}

As in most of the literature (e.g., \cite{wei_federated_2020,li2019convergence}), the convergence or accuracy error of the FL process is difficult to quantify and is analysed using the convergence bound.
Let $\bar{\mathcal{L}}^u(\bm{\sigma},{\bar{\mathbf{w}}})$ denote the upper bound of the expected global error ${\mathcal{L}}^u$ in (\ref{Equ:training accuracy loss}), then clients will take it into account when making noise-adding decisions in each iteration.
Note that the bound of the training accuracy loss function is for all FL clients, and depends on all clients' noise-adding strategies as well as the server's aggregation rule.


\textbf{Differential Privacy.}
Despite that all clients locally compute training gradient parameter $\mathbf{w}_i$ and send it to the server, attacks such as gradient inverse attack\cite{geiping2020inverting,zhao2020idlg} might disclose the original training data without accessing the datasets. 
As proposed in \cite{dwork2014algorithmic}, $(\varepsilon,\delta)$-Differential Privacy (DP) provides a criterion for privacy preservation of distributed data processing systems, and the definition is formally given as follows.
\begin{defn}[Differential Privacy]
\label{Defn:DP}
A randomized algorithm $\mathcal{P}: \mathcal{X} \rightarrow \mathcal{R}$  with domain $\mathcal{X}$ and range $\mathcal{R}$ is $(\varepsilon,\delta)$-differentially-private if for all $\mathcal{S} \subseteq \mathcal{R}$ and for all $x,y \in \mathcal{X}$ such that $||x-y||_1 \leq 1$:
\begin{equation*}
\operatorname{Pr}[\mathcal{P}(x) \in \mathcal{S}] \leq e^\varepsilon \operatorname{Pr}[\mathcal{P}(y) \in \mathcal{S}]+\delta,
\end{equation*}
where $\varepsilon$ is the distinguishable bound of all outputs on adjacent datasets $x$ and $y$ in a database, and $\delta$ represents the event that the ratio of the probabilities for two adjacent datasets $x$ and $y$ cannot be bounded by $e^\varepsilon$ after the 
 algorithm $\mathcal{P}$.
\end{defn}

Given this privacy definition, we aim to provide useful privacy protection to the transmitted parameters $\mathbf{w}_i$ and thus the raw dataset $\mathcal{D}_i$, as in \cite{wei_federated_2020}.
%
Specifically, in the $t$-th round, suppose client $i$ adds Gaussian noise $\mathbf{n}_i \sim \mathcal{N}(0,\sigma_i^2)$ locally to the transmitted parameter $\mathbf{w}_i$, as
\begin{equation*}
\tilde{\mathbf{w}}_i^{(t)}={\mathbf{w}}_i^{(t)}+{\mathbf{n}}_i^{(t)}.
\end{equation*}
According to the Gaussian mechanism\cite{dwork2014algorithmic}, the noise\footnote{For the sake of simplicity, we neglect the superscript $(t)$ for each round.} ${\mathbf{n}}_i$ with standard deviation $\sigma_{i}=\frac{c S}{\varepsilon_{i}}$
ensures $(\varepsilon_i,\delta)$-DP for any $\varepsilon_i \in (0,1)$,
where $c \geqslant \sqrt{2 \ln (1.25 / \delta)}$ is a constant, $S=\max_{\mathcal{D}_i,\mathcal{D}_i':d(\mathcal{D}_i,\mathcal{D}_i')=1} ||w_i(\mathcal{D}_i)-w_i(\mathcal{D}_i')||_2
$ is the sensitivity of the local training process.

Under the DP definition and the Gaussian mechanism, the privacy algorithm $\mathcal{P}$ outputs the noise-adding parameter $\tilde{{\mathbf{w}}}_i$ and provides a privacy guarantee to clients' transmitted parameters $\mathbf{w}_i$.
Following Definition \ref{Defn:DP}, \cite{dwork2014algorithmic} also defines a quantity to measure the privacy loss of the privacy algorithm after observing the output $\tilde{{\mathbf{w}}}_i$ as $o$, i.e.,
\begin{equation*}
\label{Equ:privacy loss}
\mathcal{L}_i^p=\ln \left( \frac{\operatorname{Pr}[\mathcal{P}({{\mathbf{w}}}_i)=o]}{\operatorname{Pr}[\mathcal{P}({{\mathbf{w}}}'_i)=o]}\right),
\end{equation*}
where ${{\mathbf{w}}}_i$ and ${{\mathbf{w}}}'_i$ are two adjacent local parameters (without noise) and $\mathcal{L}_i^p$ can be positive or negative.
%
%
Then we have
\begin{equation*}
\operatorname{Pr}(\mathcal{L}_i^p \leq  \varepsilon_i) >1- \delta.
\end{equation*} 
That is, the privacy loss $\mathcal{L}_i^p $ is upper-bounded by $\varepsilon_i$ with a large probability (given $\delta$ is small). Thus, we use $\bar{\mathcal{L}}_i^p =\varepsilon_i$ as the approximate bound of privacy loss for client $i$, to be a privacy factor affecting the client's noise-adding decision. Equivalently, we have the bound of the privacy loss 
\begin{equation}
\label{Equ:privacy loss bound}
\bar{\mathcal{L}}^p_i(\sigma_i)=\frac{c S}{\sigma_i}
\end{equation} 
from the Gaussian mechanism 
for client $i$.



\subsection{Our Mechanism Design at the Server}
\label{Subsec: server mechanism}
Given the bounds for the global training error $\bar{\mathcal{L}}^u$ and privacy loss $\bar{\mathcal{L}}_i^p$ in (\ref{Equ:privacy loss bound}), we denote $\bm{\alpha}=\{\alpha_i \in \mathcal{G}=[0,1]|i \in \mathcal{N}\}$ as the $N$ clients' privacy sensitivity types, which are private information and need to be reported by themselves to the server.
Each client $i$ minimizes a total cost function $J_i$: 
\begin{equation}
\label{Equ:loss function J_i}
	\begin{aligned}
J_i(\alpha_i,\sigma_i,\bm{\sigma}_{-i},{\bar{\mathbf{w}}})
&=
(1-\alpha_i) \bar{\mathcal{L}}^u(\bm{\sigma},{\bar{\mathbf{w}}})
+\alpha_i \bar{\mathcal{L}}_i^p(\sigma_i),
	\end{aligned}
\end{equation}
by designing the noise scale $\sigma_i$ based on their own privacy sensitivity types $\alpha_i$.
%
%
%
Note that such a type $\alpha_i$ is a constant parameter, to measure a client's individual preference on the utility-privacy tradeoff\cite{wang2020information}.

After receiving $\tilde{\mathbf{w}}_i$ with the artificial noise, the server aggregates all the local parameters as a weighted sum in (\ref{Equ:aggregation general}). 
With the aim to utilize pricing schemes for achieving better social performance at the server,
a deterministic mechanism $f$ outputs a pricing scheme $\mathbf{P}$ according to clients'  reporting $\bm{\hat{\alpha}}$ of their privacy sensitivities and local training outcomes $\tilde{\mathbf{w}}$, i.e., $f(\bm{\hat{\alpha}},\tilde{\mathbf{w}})=\{\mathbf{P}(\bm{\hat{\alpha}},\tilde{\mathbf{w}})\}$.

Considering the server's pricing scheme, each client $i$'s cost function $J_i(\alpha_i,\sigma_i,\bm{\sigma}_{-i},{\bar{\mathbf{w}}})$ in (\ref{Equ:loss function J_i}) can be modified as 
\begin{multline*}
\label{Equ:cost with pricing}
\bar{J}_i(f(\bm{\hat{\alpha}},\tilde{\mathbf{w}})|\bm{\sigma},\bm{\alpha})=
J_i(\alpha_i,\sigma_i,\bm{\sigma}_{-i},{\bar{\mathbf{w}}})
+P_i(\bm{\hat{\alpha}},\tilde{\mathbf{w}}) \\
=(1-\alpha_i) \bar{\mathcal{L}}^u(\bm{\sigma},{\bar{\mathbf{w}}})
+\alpha_i \bar{\mathcal{L}}_i^p(\sigma_i)+P_i(\bm{\hat{\alpha}},\tilde{\mathbf{w}}),
\end{multline*}
where $P_i(\bm{\hat{\alpha}},\tilde{\mathbf{w}})$ is the ex-post price charged to client $i$ after his reporting and computation.

FL clients are self-interested and cost-driven, such that they will respond to the pricing with the best noise-adding strategies to minimize their costs. Next, we explicitly model the interaction between FL clients and the server in a two-stage dynamic Bayesian game as follows.

\begin{itemize}
	\item 
	In Stage I, the server announces the number of whole training rounds or end-time $T$, the mechanism including the global aggregation rule $\bar{\textbf{w}}(\tilde{\mathbf{w}},\bm{\hat{\alpha}})$ and the (ex-post) pricing scheme $\bm{P}(\bm{\hat{\alpha}},\tilde{\mathbf{w}})$ to clients.
	The pricing schemes aim to incentivize truthful reporting of clients' types and socially optimal strategies.
	The clients' cost functions $J_i$'s in (\ref{Equ:loss function J_i}) are common knowledge.
	
	\item 
	In Stage II, clients report their types ${\hat{\alpha}_i}$'s to the server and upload parameters $\tilde{\mathbf{w}}_i$ with noise level $\sigma_i$ in distributed computation.
	Correspondingly, the server returns the global parameter ${\bar{\mathbf{w}}}$ and charges a price $P_i(\bm{\hat{\alpha}},\tilde{\mathbf{w}})$ to each client $i$
	at the end of each iteration. 
\end{itemize}

Given that clients' privacy sensitivity $\alpha_i$'s are private information, for achieving  better social performance, the mechanism design at the server needs to be truthful and efficient.
%
Next we formally define the truthfulness.
\begin{defn}[Truthfulness]
A mechanism is truthful if no agent can benefit from misreporting his type $i$. Formally, given client $i$, type profile $\bm{\alpha}=\{\alpha_i,\bm{\alpha}_{-i}\} \in \mathcal{G}^N$, and any misreported type ${\hat{\alpha}_i} \in \mathcal{G}$, it holds that 
\begin{eqnarray*}
\bar{J}_i(f(\alpha_i,\bm{\alpha}_{-i},\tilde{\mathbf{w}})|\bm{\sigma},\bm{\alpha}) \leq
\bar{J}_i(f({\hat{\alpha}_i},\bm{\alpha}_{-i},\tilde{\mathbf{w}})|\bm{\sigma},\bm{\alpha}).
\end{eqnarray*}
\end{defn}

We are also interested in designing truthful mechanisms that perform well with respect to minimizing the social cost. Define
the social cost of a mechanism $f(\bm{\hat{\alpha}},\tilde{\mathbf{w}})$ as the sum of costs of $N$ agents, i.e.,
\begin{equation}
\label{Equ:SC}
\begin{aligned}
SC(f(\bm{\hat{\alpha}},\tilde{\mathbf{w}})|\bm{\sigma},\bm{\alpha})
&=\sum_{i \in \mathcal{N}}J_i(\alpha_i,\bm{\sigma},{\bar{\mathbf{w}}}).
\end{aligned}
\end{equation}
Let $OPT(\bm{\alpha})$ denote the optimal minimum social cost, i.e.,
\begin{equation*}
\label{Equ:OPT social opt}
OPT(\bm{\alpha})=\min_{\bm{\sigma}} SC(f(\bm{\hat{\alpha}},\tilde{\mathbf{w}})|\bm{\sigma},\bm{\alpha}).
\end{equation*}
Then we use the concept of price of anarchy (PoA) to tell the maximum efficiency loss due to clients' selfish behavior by adding too much noise or misreporting types.

\begin{defn}[Price of Anarchy]
\label{Def:POA}
The ratio between the social cost and the socially optimal cost for mechanism $f(\bm{\hat{\alpha}},\tilde{\mathbf{w}})$ is 
\begin{equation*}
\gamma=\frac{SC(f(\bm{\hat{\alpha}},\tilde{\mathbf{w}})|\bm{\sigma},\bm{\alpha})}{OPT(\bm{\alpha})} \geq 1.
\end{equation*}
The Price of Anarchy (PoA) is defined as the maximum of the ratio by scanning through all possible parameters:
\begin{equation*}
PoA=\max_{\bm{\alpha},N, c,S,T,\lambda_F,L_F} \gamma.
\end{equation*}
\end{defn}

The PoA definition describes an efficiency gap between clients' equilibrium strategy and the social optimum. If it is non-small, it motivates us to design a mechanism which approaches or even achieves social optimum with $\gamma=1$.
Next, we will analyse clients' equilibrium strategy and prepare the server's mechanism design. 

\section{Game Analysis and Mechanism Design under Complete information}
\label{Sec:compl- infor}
In this section, we assume that all clients' types are known to the parameter server (or equivalently, ${\hat{\alpha}_i}=\alpha_i$), which may be done by checking their historical activities on similar FL tasks. In this scenario, the server's mechanism $f(\bm{\hat{\alpha}},\tilde{\mathbf{w}})$ reduces to a function of parameters $\tilde{\mathbf{w}}$.
We derive clients' equilibrium noise-adding strategy after introducing a properly improved aggregation rule at the server, and then we design a mechanism to achieve the minimum social cost. 

\subsection{Improved Aggregation Rule at the Server}
\label{Subsec:aggregation rule at the server}
Given the clients' training accuracy loss $\bar{\mathcal{L}}^u(\bm{\sigma},{\bar{\mathbf{w}}})$ in (\ref{Equ:loss function J_i}) depends on the server's aggregation rule, we first consider an aggregation rule ${\bar{\mathbf{w}}}(\bm{\hat{\alpha}},\tilde{\mathbf{w}})={\bar{\mathbf{w}}}(\bm{\alpha},\tilde{\mathbf{w}})$ from (\ref{Equ:aggregation general}) to ensure the training convergence performance.

%

The standard aggregation rule in the FedAvg algorithm \cite{li2019convergence} uses a simple mean aggregation as
\begin{equation}
\label{Equ:mean aggregation}
\bar{\mathbf{w}}^{mean}(\bm{{\alpha}},\tilde{\mathbf{w}})=\frac{\sum_{i \in \mathcal{N}}  \tilde{\mathbf{w}}_{i}}{N},
\end{equation}
which does not take clients' strategic behaviours or added noise variances into consideration.
Alternatively, predicting any client $i$' strategy with distinct standard deviation $\sigma_i(\alpha_i)$ from cost function (\ref{Equ:loss function J_i}), we properly use the maximum likelihood estimator (MLE) for better aggregation with $\bm{\hat{\alpha}}=\bm{\alpha}$:
\begin{equation}
\label{Equ:MLE}
{\bar{\mathbf{w}}}^{MLE}(\bm{{\alpha}},\tilde{\mathbf{w}})=\frac{\sum_{i \in \mathcal{N}} \frac{\tilde{\mathbf{w}}_{i}}{\sigma_{i}^{2}(\alpha_i)}}{\sum_{i \in \mathcal{N}} \frac{1}{\sigma_{i}^{2}(\alpha_i)}}.
\end{equation}

\begin{prop}
\label{Prop:bound of training accuracy loss with MLE}
With the MLE aggregation rule in (\ref{Equ:MLE}), we give the upper bound of the training accuracy loss function ${\mathcal{L}}^u$ in (\ref{Equ:training accuracy loss}) as
\begin{equation}
\label{Equ:L_u MLE}
\bar{\mathcal{L}}^u(\bm{\sigma},{\bar{\mathbf{w}}})=\kappa 
\Delta^{MLE}(\bm{\sigma})
\left(1+\frac{1}{2 L_{F}} \Delta^{MLE}(\bm{\sigma})\right),
\end{equation}
where $\kappa=16 ||\mathbf{w}^0-\mathbf{w}^*||_2$ with $T=\frac{L_F}{\Delta}||\mathbf{w}^0-\mathbf{w}^*||_2$, step size $\eta=1/L_F$, $\Delta^{MLE}=(\sum_{i=1}^{N} \sigma_{i}^{-2})^{-\frac{1}{2}}$ measures the bound of the expected error $\mathbb{E}|\hat{\mu}-\mu|$ achieved by MLE.
Our improved aggregation rule in (\ref{Equ:MLE}) leads to a smaller convergence bound than the mean aggregation in (\ref{Equ:mean aggregation}).
\end{prop}

Proposition \ref{Prop:bound of training accuracy loss with MLE} guides us to choose the MLE aggregation rule for better convergence performance at the server.

%

\subsection{Clients' Strategy under Equilibrium and Social Optimum}
\label{Subsubsec:user NE with MLE aggregation rule}
By substituting (\ref{Equ:L_u MLE}) into the cost function (without pricing) for client $i$ in (\ref{Equ:loss function J_i}), the cost for client $i$ becomes  
\begin{equation*}
	\begin{aligned}
J_i(\sigma_i,\bm{\sigma}_{-i})
&=
\kappa (1-\alpha_i)
\Delta^{MLE}(\bm{\sigma})
(1+\frac{\Delta^{MLE}(\bm{\sigma})}{2 L_{F}} )
+ \frac{\alpha_i c S }{\sigma_i}.
\end{aligned}
\end{equation*}
By minimizing the individual's and social costs, we give the equilibrium and social optimal noise-adding strategies for client $i$.

\begin{prop}
Given that the server chooses the MLE aggregation rule in (\ref{Equ:MLE}), the equilibrium noise level $\sigma_i^{*}$ for client $i$ with privacy sensitivity type $\alpha_i$ is the unique solution to
\begin{align}
\label{Equ:sigma NE}
\kappa (1-\alpha_i) 
(\sum_{j \in \mathcal{N}} \sigma_{j}^{-2})^{-\frac{3}{2}}
 \frac{
\left(\sum_{j \in \mathcal{N}} \sigma_{j}^{-2}\right)^{-\frac{1}{2}}
+L_F}{L_F}
=\alpha_i c S \sigma_i,
\end{align}
while under social optimum, the noise level $\sigma_i^{**}$ for client $i$ 
is the unique solution to
\begin{multline}
\label{Equ:sigma SO}
\kappa \left( 
\sum_{i \in \mathcal{N}} (1-\alpha_i)
\right) 
(\sum_{j \in \mathcal{N}} \sigma_{j}^{-2})^{-\frac{3}{2}}
 \frac{
\left(\sum_{j \in \mathcal{N}} \sigma_{j}^{-2}\right)^{-\frac{1}{2}}
+L_F}{L_F} \\
=\alpha_i c S \sigma_i.
\end{multline}
\end{prop}

To show the maximum gap of social costs defined in (\ref{Equ:SC}) under clients' equilibrium in (\ref{Equ:sigma NE}) and the social optimum in (\ref{Equ:sigma SO}), we use the PoA definition in Definition \ref{Def:POA}.

\begin{prop}
\label{Prop:PoA}
Price of anarchy under complete information goes to infinity under the worst case.
\end{prop}


\subsection{Pricing Mechanism for Achieving Social Optimum}

To incentivize client $i$ to choose noise level $\sigma_i^{**}$ rather than $\sigma_i^{*}$, we design the ex-post pricing scheme $\mathbf{P}(\bm{\hat{\alpha}},\tilde{\mathbf{w}})=\mathbf{P}(\bm{\alpha},\tilde{\mathbf{w}})$ to impose penalty to clients who add too much noise to their local parameters $\mathbf{w}_i$'s.



\begin{defn}[Pricing Mechanism under Complete Information]
\label{Def:mechanism CI}
Under complete information of clients' types $\bm{\alpha}$, the mechanism $f({\bm{\alpha}},\tilde{\mathbf{w}})=\{\mathbf{P}(\bm{\alpha},\tilde{\mathbf{w}})\}$ decides pricing function
\begin{align*}
P_i(\bm{\alpha},\tilde{\mathbf{w}})
&=
\beta_i^*(\bm{\alpha})\left( \tilde{\mathbf{w}}_i^{}-\frac{1}{N}\sum_{i=1}^N \tilde{\mathbf{w}}_i\right) ^2
-q,
\end{align*}
charged to client $i$,
where the penalty coefficient is
\begin{equation}
\label{Equ:penalty coefficient beta under CI}
\beta_i^*(\bm{\alpha})
=\frac{N^2 \sum_{j \neq  i,j\in \mathcal{N}} (1-\alpha_j) \alpha_j  c S}
{2(N-1)^2 (\sum_{i\in \mathcal{N}} (1-\alpha_i) (\sigma_i^{**})^3)},
\end{equation}
and the compensation payment is
\begin{align*}
q=
\frac{2(N-1)^2}{N} \sum_{i \in \mathcal{N}}
\beta_i^*(\bm{\alpha}) \sigma_i^{**},
\end{align*}
where $\sigma_i^{**}$ is given in (\ref{Equ:sigma SO}). 
\end{defn}

\begin{prop}
\label{Prop:property CI}
Under complete information, our pricing mechanism in Definition \ref{Def:mechanism CI} simultaneously reaches the social optimum with $\gamma=1$ and ex-ante budget balance.
\end{prop}

Though our pricing scheme with (\ref{Equ:penalty coefficient beta under CI}) does not verify each client's noise/contribution and simply compares each client's performance to the noisy average benchmark, it already achieves the social optimum. Even when all clients have the same sensitivity types, they worry their randomly computed  parameters to be an outlier and do not dare to add much noise. Ex-ante budget balance for each client (achieved by compensation $q$) is an important property in economics\cite{courcoubetis2009economic} and  helps ensure long-term client participation.  

\begin{figure}[!t]
\centering
\includegraphics[width=2 in]{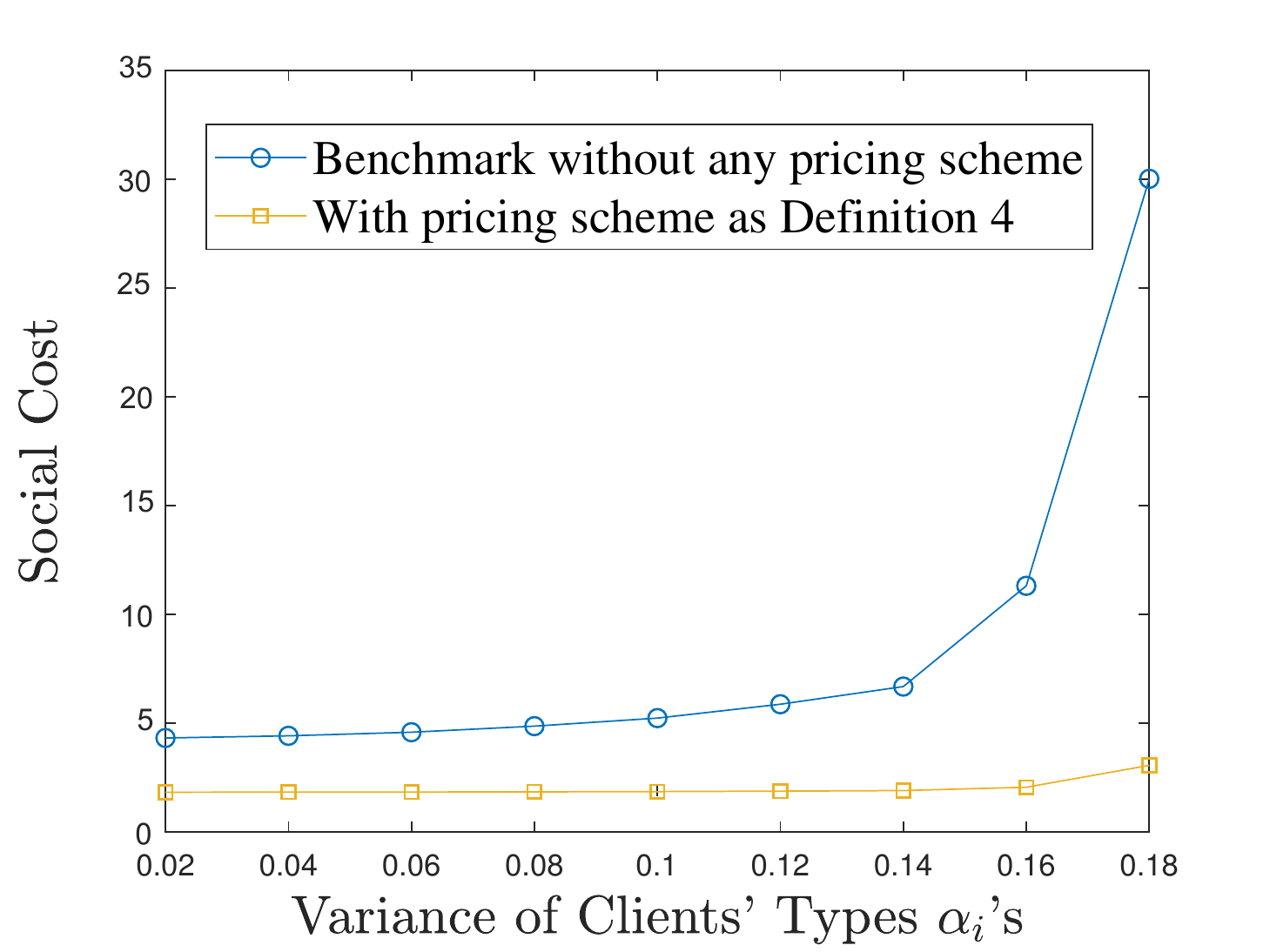}
\caption{Social costs under complete information with or without our pricing mechanism in Definition \ref{Def:mechanism CI} versus the variance of clients' types $\alpha_i$. Here we set $\alpha_i$ to follow a normal distribution with mean $0.5$.
}
\label{Fig:div 100 user}
\end{figure}

Without a pricing scheme, Fig.~\ref{Fig:div 100 user} shows a substantial increase in social cost as the variance of $\alpha_i$ rises.
This trend can be attributed to the increasingly diverse objectives of varying clients, leading to differences in their training goals and privacy concerns.
Self-interested client behavior to protect privacy leads to more noise and higher training accuracy loss. However, our pricing scheme maintains lower social costs due to clients' sensitivity to penalties.

\section{Game Analysis and Mechanism Design  under Incomplete information}
\label{Sec:Incomplete information}

Under incomplete information, all clients' real types $\bm{\alpha}$ are private information while the only common knowledge 
is the random type distribution $\alpha_i \sim \mathcal{G}(\alpha)$ in $[0,1]$. We propose that the server asks each client $i$ to report his type ${\hat{\alpha}_i}$, which might not be truthful due to clients' selfish behaviours.
The Bayesian game for server-client interaction is formulated as in Section \ref{Subsec: server mechanism}.
We refine our pricing scheme in Definition \ref{Def:mechanism CI} to fit. 



\begin{defn}[Pricing Mechanism under Incomplete Information]
\label{Def:mechanism II}
Under incomplete information of clients' types $\bm{\alpha}$, our mechanism $f(\bm{\hat{\alpha}},\tilde{\mathbf{w}})=\{\mathbf{P}(\bm{\hat{\alpha}},\tilde{\mathbf{w}})\}$ decides:
\begin{align*}
P_i(\bm{\hat{\alpha}},\tilde{\mathbf{w}})
&=
{p}_i(\bm{\hat{\alpha}},\tilde{\mathbf{w}})-r_i(\bm{\hat{\alpha}})
-q.
\end{align*}
\begin{itemize}
\item 
The penalty $p_i(\bm{\hat{\alpha}},\tilde{\mathbf{w}})$  aims to incentivize client $i$ to update local parameters with only socially optimal noise variance ${\sigma}_i^{**}$.
It takes a similar form from Definition \ref{Def:mechanism CI} as $p_i(\bm{\hat{\alpha}},\tilde{\mathbf{w}})=
\beta_i(\bm{\hat{\alpha}})\left( \tilde{\mathbf{w}}_i^{}-\frac{1}{N}\sum_{i=1}^N \tilde{\mathbf{w}}_i\right) ^2$, yet here we need to design new parameters $\bm{\beta}(\bm{\hat{\alpha}})$ under incomplete information.

\item 
The reward function $r_i(\bm{\hat{\alpha}})$ is to incentivize client $i$'s truthful reporting of $\alpha_i$ under incomplete information. 
\item 
The constant compensation payment $q$ is refunded to each client to make the FL system ex-ante budget balanced. 
\end{itemize}
\end{defn}


Next, we analyse the Bayesian game between clients and the server under incomplete information by backward induction.

\subsection{Each Client's Decision of ${\hat{\alpha}_i}$ and $\sigma_i$ in Stage II}
Given the pricing mechanism $\mathbf{P}=(\bm{r}(\bm{\hat{\alpha}}),\bm{p}(\bm{\hat{\alpha}},\tilde{\mathbf{w}}),q)$ announced in Stage I from Definition \ref{Def:mechanism II}, client $i$'s strategy based on his own private type $\alpha_i$ includes two parts: the type reporting strategy ${\hat{\alpha}_i}(\alpha_i)$ and the noise variance $\sigma_i(\alpha_i)$.
%

\textbf{Aggregation Rule under Incomplete Information.}
Similar to the aggregation rule under complete information in Section \ref{Subsec:aggregation rule at the server},
we still use the MLE aggregation rule in (\ref{Equ:MLE}), yet we have to replace real $\sigma_i$ with $\sigma_{i}^*(\alpha_i={\hat{\alpha}_i})$ according to clients' reported type ${\hat{\alpha}_i}$.
In this case, we modify the training accuracy loss function $\bar{\mathcal{L}}^u(\bm{\sigma},{\bar{\mathbf{w}}})$ in (\ref{Equ:L_u MLE}) 
with 
$$\Delta^{inc}(\bm{\sigma}^*,\bm{\hat{\alpha}})=
\frac{
\left(\sum_{i=1}^N 
\frac{(\sigma_i^{*}(\alpha_i))^2}{(\sigma_i^{*}(\alpha_i={\hat{\alpha}_i}))^4}
\right)^{\frac{1}{2}}
}{
\sum_{i=1}^N \frac{1}{(\sigma_i^{*}(\alpha_i={\hat{\alpha}_i}))^2}},$$ which depends on clients' reported types $\bm{\hat{\alpha}}$ and actual noise strategies $\bm{\sigma}^{*}(\bm{\alpha})$ at the Bayesian Nash equilibrium (BNE).

\textbf{Client $i$'s Expected Cost.}
%
%
Given all clients' types follow an i.i.d. distribution $\mathcal{G}$, client $i$ needs to estimate all other $N-1$ clients' strategies $\bm{\hat{\alpha}}_{-i}(\bm{\alpha}_{-i})$ and $\bm{\sigma}_{-i}(\bm{\alpha}_{-i})$ with $\alpha_j \sim \mathcal{G}(\alpha)$ $(j \in \mathcal{N}/i)$ when making his own strategies.
%
%
%
With the pricing scheme $P_i(\bm{\hat{\alpha}},\tilde{\mathbf{w}})$ in Definition \ref{Def:mechanism II}, we take expectation over $\bm{\alpha}_{-i}$ to obtain the expected cost function for client $i$:
\begin{equation}
\begin{aligned}
\label{Equ: expected cost under incom- infor-}
&\mathbb{E}_{\bm{\alpha}_{-i}}J_i(\alpha_i,{\hat{\alpha}_i},\sigma_i|r_i(\bm{\hat{\alpha}}),p_i(\bm{\hat{\alpha}},\tilde{\mathbf{w}}),q) \\=
&(1-\alpha_i) \mathbb{E}_{\bm{\alpha}_{-i}}\bar{\mathcal{L}}^u
({\hat{\alpha}_i},\sigma_i,\bm{\hat{\alpha}}_{-i}(\bm{\alpha}_{-i}),\bm{\sigma}_{-i}(\bm{\alpha}_{-i})) 
+\alpha_i\bar{\mathcal{L}}_i^p(\sigma_i) 
\\
-&\mathbb{E}_{\bm{\alpha}_{-i}}r_i(\bm{\hat{\alpha}})
-q
\\
+&
\mathbb{E}_{\bm{\alpha}_{-i}}
\left[ 
\beta_i(\bm{\hat{\alpha}})\left( (\frac{N-1}{N})^2 \sigma_i^2+\frac{\sum_{j=1,j\neq i}^N(\sigma_j(\alpha_j))^2}{N^2}\right)
\right] .
\end{aligned}
\end{equation}

%
%
%


Given the pricing mechanism $\mathbf{P}=(\bm{r}(\bm{\hat{\alpha}}),\bm{p}(\bm{\hat{\alpha}},\tilde{\mathbf{w}}),q)$ announced in Stage I, client $i$ will optimize noise level $\sigma_i^*$ as the solution to $\frac{\partial \mathbb{E}_{\bm{\alpha}_{-i}}J_i(\alpha_i,{\hat{\alpha}_i},\sigma_i)}{\partial \sigma_i}=0$
and $\hat{\alpha}_i^{*}$ as the solution to
$\frac{\partial \mathbb{E}_{\bm{\alpha}_{-i}}J_i(\alpha_i,{\hat{\alpha}_i},\sigma_i)}{\partial {\hat{\alpha}_i}}=0$ at the BNE to minimize 
$\mathbb{E}_{\bm{\alpha}_{-i}}J_i$ in (\ref{Equ: expected cost under incom- infor-}).
Next we only focus on the symmetric BNE strategy for each client due to the symmetry of (\ref{Equ: expected cost under incom- infor-}), i.e., the noise-adding strategy $\sigma(\alpha_i)$ and the type reporting strategy $\hat{\alpha}(\alpha_i)$ without subscripts in these two functions.

\subsection{Clients' Equilibrium Strategies with Binary Type Values}
\label{Subsec: binary case}
To simplify the analysis, we limit clients' i.i.d. private type distribution of $\alpha_i$ to be two discrete values in $\{\alpha_L,\alpha_H\}$ with $0\leq \alpha_L <\alpha_H \leq 1$, though similar analysis methods holds for the continuous distribution. The common knowledge is the proportion $\eta$ of all clients' types:
\begin{equation*}
\alpha_i=
\begin{cases}
\alpha_L, &\text{ with probability } \eta,\\
\alpha_H, &\text{ with probability } 1-\eta.
\end{cases}
\end{equation*}

Considering the deterministic reporting strategy ${\hat{\alpha}_i} \in \{\alpha_L,\alpha_H\}$, clients may 
truthfully report or misreport.
Due to the symmetry of both types of clients, there are three cases for all clients' type reporting strategies $\bm{\hat{\alpha}}$ in general.

\subsubsection{Case 1}
all clients reports the same type $\alpha_L$ or $\alpha_H$, i.e., $\hat{\alpha}(\alpha_L)=\hat{\alpha}(\alpha_H) \in \{\alpha_L,\alpha_H\}$ for any $i \in \mathcal{N}$.

In this case, receiving the same type reporting from all clients, the server has to assign the same weight with $\rho_i=1$  in the aggregation rule in (\ref{Equ:aggregation general}).
%
Also, the reward function and penalty coefficient are indistinctive among all agents.

\begin{lemma}
\label{Lemma:noise under case 1}
In Case 1 where all clients report the same type $\hat{\alpha}(\alpha_L)=\hat{\alpha}(\alpha_H) \in \{\alpha_L,\alpha_H\}$ to the server, all clients' noise adding strategies 
$\sigma^{case1}(\alpha_L)$ and $\sigma^{case1}(\alpha_H)$ at the BNE are solutions to the equation set:
{\tiny
\begin{equation*}
\begin{cases}
(1-\alpha_L) \kappa
\left[ 
\frac{\sigma(\alpha_L)}{N}
\sum_{n=0}^{N-1}
\left\lbrace 
p(n)
((n+1)\sigma^2(\alpha_L)
+(N-1-n) \sigma^2(\alpha_H)
)^{-1/2}\right\rbrace 
+ 
\right. \\
\left.
\frac{\sigma(\alpha_L)}{N^2 L_F} \right]
+2\beta(\frac{N-1}{N})^2 \sigma(\alpha_L)
=\alpha_L\frac{cS}{\sigma^2(\alpha_L)},\\
(1-\alpha_H) \kappa
\left[ 
\frac{\sigma(\alpha_H)}{N}
\sum_{n=0}^{N-1}
\left\lbrace 
p(n)
(n\sigma^2(\alpha_L)
+(N-n) \sigma^2(\alpha_H)
)^{-1/2}\right\rbrace 
+
\frac{\sigma(\alpha_H)}{N^2 L_F} \right]
\\
+2\beta(\frac{N-1}{N})^2 \sigma(\alpha_H)
=\alpha_H\frac{cS}{\sigma^2(\alpha_H)},
\end{cases}
\end{equation*}
}
where $\beta=\beta^L$ if $\hat{\alpha}(\alpha_L)=\hat{\alpha}(\alpha_L)=\alpha_L$, and $\beta=\beta^H$ if $\hat{\alpha}(\alpha_L)=\hat{\alpha}(\alpha_L)=\alpha_H$, $p(n)=\operatorname{Pr}(N_L=n)=C_{N-1}^{n}\eta^n(1-\eta)^{N-1-n}$.

\end{lemma}

\subsubsection{Case 2}
All clients misreport their private types to the server, i.e., ${\hat{\alpha}}(\alpha_L)=\alpha_H$ and ${\hat{\alpha}}(\alpha_H)=\alpha_L$. 
In this case, the server will use the aggregation rule (\ref{Equ:MLE}) with $\sigma_{i}^*(\alpha_i={\hat{\alpha}_i})$ according to clients' misreported type ${\hat{\alpha}_i}$.

\begin{lemma}
\label{Lemma:noise under case 2}
In Case 2 where all clients misreport their types to the server, all clients' noise adding strategies $\sigma^{case2}(\alpha_L)$ and $\sigma^{case2}(\alpha_H)$ at the BNE are  solutions to the equation set:
{\tiny
\begin{equation*}
\begin{cases}
(1-\alpha_L) \kappa
\frac{\sigma(\alpha_L)}{\sigma^4({\alpha_H})}
\sum_{n=0}^{N-1}
p(n)
\left[ 
\frac{
\left(
(n+1)
\frac{\sigma^2(\alpha_L)}{\sigma^4(\alpha_H)}
+
(N-1-n)
\frac{\sigma^2(\alpha_H)}{\sigma^4(\alpha_L)}
\right)^{-\frac{1}{2}}
}{
(n+1)
\frac{1}{\sigma^2(\alpha_H)}
+
(N-1-n)
\frac{1}{\sigma^2(\alpha_L)}} \right. \\
+
\left.
\frac{
1
}{
L_F((n+1)
\frac{1}{\sigma^2(\alpha_H)}
+
(N-1-n)
\frac{1}{\sigma^2(\alpha_L)})^2}
\right]
+2\beta^H(\frac{N-1}{N})^2 \sigma(\alpha_L)
=\alpha_L\frac{cS}{\sigma^2(\alpha_L)},\\
(1-\alpha_H) \kappa
\frac{\sigma(\alpha_H)}{\sigma^4({\alpha_L})}
\sum_{n=0}^{N-1}
p(n)
\left[ 
\frac{
\left(
n
\frac{\sigma^2(\alpha_L)}{\sigma^4(\alpha_H)}
+
(N-n)
\frac{\sigma^2(\alpha_H)}{\sigma^4(\alpha_L)}
\right)^{-\frac{1}{2}}
}{
n
\frac{1}{\sigma^2(\alpha_H)}
+
(N-n)
\frac{1}{\sigma^2(\alpha_L)}}
\right. 
\\
\left.
+
\frac{
1
}{
L_F(n
\frac{1}{\sigma^2(\alpha_H)}
+
(N-n)
\frac{1}{\sigma^2(\alpha_L)})^2}
\right]
+2\beta^L(\frac{N-1}{N})^2 \sigma(\alpha_H)
=\alpha_H\frac{cS}{\sigma^2(\alpha_H)}.
\end{cases}
\end{equation*}
}

\end{lemma}

\subsubsection{Case 3}
All clients report their types truthfully, i.e., ${\hat{\alpha}}(\alpha_L)=\alpha_L$, ${\hat{\alpha}}(\alpha_H)=\alpha_H$. 
In this case, the server will use the aggregation rule (\ref{Equ:MLE}) with $\sigma_{i}^*(\alpha_i={\hat{\alpha}_i})$ according to clients' reported type ${\hat{\alpha}_i}$.
Similarly, the reward function and penalty coefficient $\bm{\beta}(\bm{\hat{\alpha}})$ include two parts, depending on the outcome type ${\hat{\alpha}_i}$.

\begin{lemma}
\label{Lemma:noise under case 3}
In Case 3 where all clients reports their types ${\hat{\alpha}_L}(\alpha_L)=\alpha_L$ and ${\hat{\alpha}_H}(\alpha_H)=\alpha_H$ truthfully to the server, all clients' noise adding strategies $\sigma^{case3}(\alpha_L)$ and $\sigma^{case3}(\alpha_H)$ at the BNE are  solutions to the equation set:
{\tiny
\begin{equation*}
\begin{cases}
(1-\alpha_L) \kappa
\sigma^{-3}(\alpha_L)
\sum_{n=0}^{N-1}
p(n)
\left[ 
(1+\frac{\left(
(n+1)
\sigma^{-2}(\alpha_L)
+
(N-1-n)
\sigma^{-2}(\alpha_H)\right)^{-\frac{1}{2}}}{L_F}) 
\right. \\
\left.
\left(
(n+1)
\sigma^{-2}(\alpha_L)
+
(N-1-n)
\sigma^{-2}(\alpha_H)\right)^{-\frac{3}{2}}
\right]
\\ 
+2\beta^L(\frac{N-1}{N})^2 \sigma(\alpha_L)
=\alpha_L\frac{cS}{\sigma^2(\alpha_L)},
\\
(1-\alpha_H) \kappa
\sigma^{-3}(\alpha_H)
\sum_{n=0}^{N-1}
p(n)
\left[ 
(1+\frac{\left(
n
\sigma^{-2}(\alpha_L)
+
(N-n)
\sigma^{-2}(\alpha_H)\right)^{-\frac{1}{2}}}{L_F})
\right. \\
\left.
\left(
n
\sigma^{-2}(\alpha_L)
+
(N-n)
\sigma^{-2}(\alpha_H)\right)^{-\frac{3}{2}}
\right]
+2\beta^H(\frac{N-1}{N})^2 \sigma(\alpha_H)
=\alpha_H\frac{cS}{\sigma^2(\alpha_H)}.
\end{cases}
\end{equation*}
}

\end{lemma}

Notice that in all cases, the server imposes penalties and rewards predicated on all clients' reported types $\bm{\hat{\alpha}}$. For example, we give the penalty coefficient $\beta_i(\hat{\alpha}_i)$ as 
\begin{equation*}
\beta_i(\hat{\alpha}_i)=
\begin{cases}
\beta^L, &\text{ if } {\hat{\alpha}_i}=\alpha_L, \\
\beta^H, &\text{ if } {\hat{\alpha}_i}=\alpha_H.
\end{cases}
\end{equation*}  
Both noise-adding strategies $\sigma(\alpha_L)$ and $\sigma(\alpha_H)$ are functions of clients' actual type  $\alpha_i \in \{\alpha_L,\alpha_H \}$, and depends on clients' type distribution parameter $\eta$, total number of clients $N$, and the penalty coefficients $\bm{\beta}$.
By comparing 
the expected cost for the minimum, each client $i$ will choose the type reporting $\hat{\alpha}_i$ at the BNE from the three cases above.

\subsection{Pricing Mechanism Design in Stage I}
After analysing all clients' type reporting and noise adding strategies in Stage II, in this subsection we are ready to proactively design the pricing mechanism $\mathbf{P}=(\bm{r}(\bm{\hat{\alpha}}),\bm{p}(\bm{\hat{\alpha}},\tilde{\mathbf{w}}),q)$ from Definition \ref{Def:mechanism II} in Stage I, with the goal of truthful reporting, social efficiency and ex-ante budget balance.

First we define some notations to simplify the BNE results: let 
$Q_1=\sum_{n=0}^{N-1}
C_{N-1}^n \eta^n(1-\eta)^{N-1-n}
\cdot
p(N-1-n)$,
$Q_2=\sum_{n=0}^{N-1}
C_{N-1}^n \eta^n(1-\eta)^{N-1-n}
\cdot
p(n+1)$,
$$X_1(n)
=(n+1)(\sigma^{**}(\alpha_L))^{-2}+(N-1-n) (\sigma^{**}(\alpha_H))^{-2},$$
$$X_2(n)
=n(\sigma^{**}(\alpha_L))^{-2}+(N-n) (\sigma^{**}(\alpha_H))^{-2},$$
where
$(\sigma^{**}(\alpha_L),\sigma^{**}(\alpha_H))$ is the social optimal noise level,
$p(n)=C_N^{n} \eta^{n} (1-\eta)^{N-n}$ denote the probability that the true number of clients with type $\alpha_L$ is exactly $n$.

\begin{prop}
\label{Prop:pricing scheme}
In the pricing mechanism $\mathbf{P}=(\bm{r}(\bm{\hat{\alpha}}),\bm{p}(\bm{\hat{\alpha}},\tilde{\mathbf{w}}),q)$ in Definition \ref{Def:mechanism II}, by setting the penalty coefficient $\bm{\beta}(\bm{\hat{\alpha}})$ with
{\tiny
\begin{multline*}
\beta^L=
\frac{\kappa N^2}{2(N-1)^2 (\sigma^{**}(\alpha_L))^4}
\left\lbrace 
(1-\alpha_L) 
\sum_{n=0}^{N-1}
\left[ 
p(n)
n
\left( 
1+\frac{X_1^{-\frac{1}{2}}(n)}{L_F}
\right)  
X_1^{-\frac{3}{2}}(n)
\right]  
\right.
\\
+
\left.
\frac{1-\eta}{\eta}  
(1-\alpha_H) 
\sum_{n=0}^{N-1}
\left[ 
p(n)
n
\left( 
1+\frac{X_2^{-\frac{1}{2}}(n)}{L_F}
\right)  
X_2^{-\frac{3}{2}}(n)
\right] 
\right\rbrace,
\end{multline*}
}
{\tiny
\begin{multline*}
\beta^H= 
\frac{\kappa N^2}{2(N-1)^2 (\sigma^{**}(\alpha_H))^4} \\
\left\lbrace 
\frac{\eta}{1-\eta}
(1-\alpha_L)
\sum_{n=0}^{N-1}
\left[ 
p(n)
(N-1-n)
\left( 
1+\frac{X_1^{-\frac{1}{2}}(n)}{L_F}
\right)  
X_1^{-\frac{3}{2}}(n)
\right]  
\right.
\\
+
\left.
(1-\alpha_H) 
\sum_{n=0}^{N-1}
\left[ 
p(n) (N-n)
\left( 
1+\frac{X_2^{-\frac{1}{2}}(n)}{L_F}
\right)  
X_2^{-\frac{3}{2}}(n)
\right] 
\right\rbrace,
\end{multline*}
}
and the reward function
\begin{equation*}
r_i(\hat{\bm{\alpha}})
=
\begin{cases}
r^L p({\hat{N}_L}(\hat{\bm{\alpha}})), &\text{ if } \hat{\alpha}_i=\alpha_L, \\
r^H p({\hat{N}_L}(\hat{\bm{\alpha}})), &\text{ if } \hat{\alpha}_i=\alpha_H,
\end{cases}
\end{equation*}
with 
$r^L=\max\left( \frac{Q_2(A_3-A_1)-Q_1(A_2-A_4)}{Q_2^2-Q_1^2},0\right)$, 
$r^H=
\max\left(
\frac{Q_1(A_3-A_1)-Q_2(A_2-A_4)}{Q_2^2-Q_1^2},0\right)$,
and ${\hat{N}_L}(\hat{\bm{\alpha}})=\sum_{i\in \mathcal{N}}1_{\hat{\alpha}_i=\alpha_L}$ denoting the number of received reporting types of $\alpha_L$,
our pricing mechanism ensures all clients' truthful type reporting and achieves the social optimum of noise-adding. The detailed forms of $A_i$ (i=1,2,3,4) are given in the on-line technical report \cite{TechnicalReport}. 
\end{prop}

Notice that $p({\hat{N}_L}(\hat{\bm{\alpha}}))$ denote the probability that the real number of clients with type $\alpha_L$ is exactly the number ${\hat{N}_L}(\hat{\bm{\alpha}})$ of received number of reporting with type $\alpha_L$. This probability guarantees the incorporation of all reporting decisions into the reward function, thereby preventing clients from deviating. 
For instance, one cannot misreport for a higher reward, as such symmetric misreporting by other clients of the same type would result in a reduced reward for all.
The penalty coefficients are designed through a comparison of clients' equilibrium and socially optimal strategies, as seen in complete information scenarios.
We 
also ensure the ex-ante budget balance by numerically designing the 
compensation payment in Definition \ref{Def:mechanism II} as:
$$q
=\frac{\mathbb{E}_{\bm{\alpha} \sim \mathcal{G}^N} \left( \sum_{i \in \mathcal{N}}p_i(\bm{\hat{\alpha}}(\bm{\alpha}))
-
\sum_{i \in \mathcal{N}}r_i(\bm{\hat{\alpha}}(\bm{\alpha}))
\right) }{N}.
$$

\begin{figure}[!t]
\centering
\includegraphics[width=2 in]{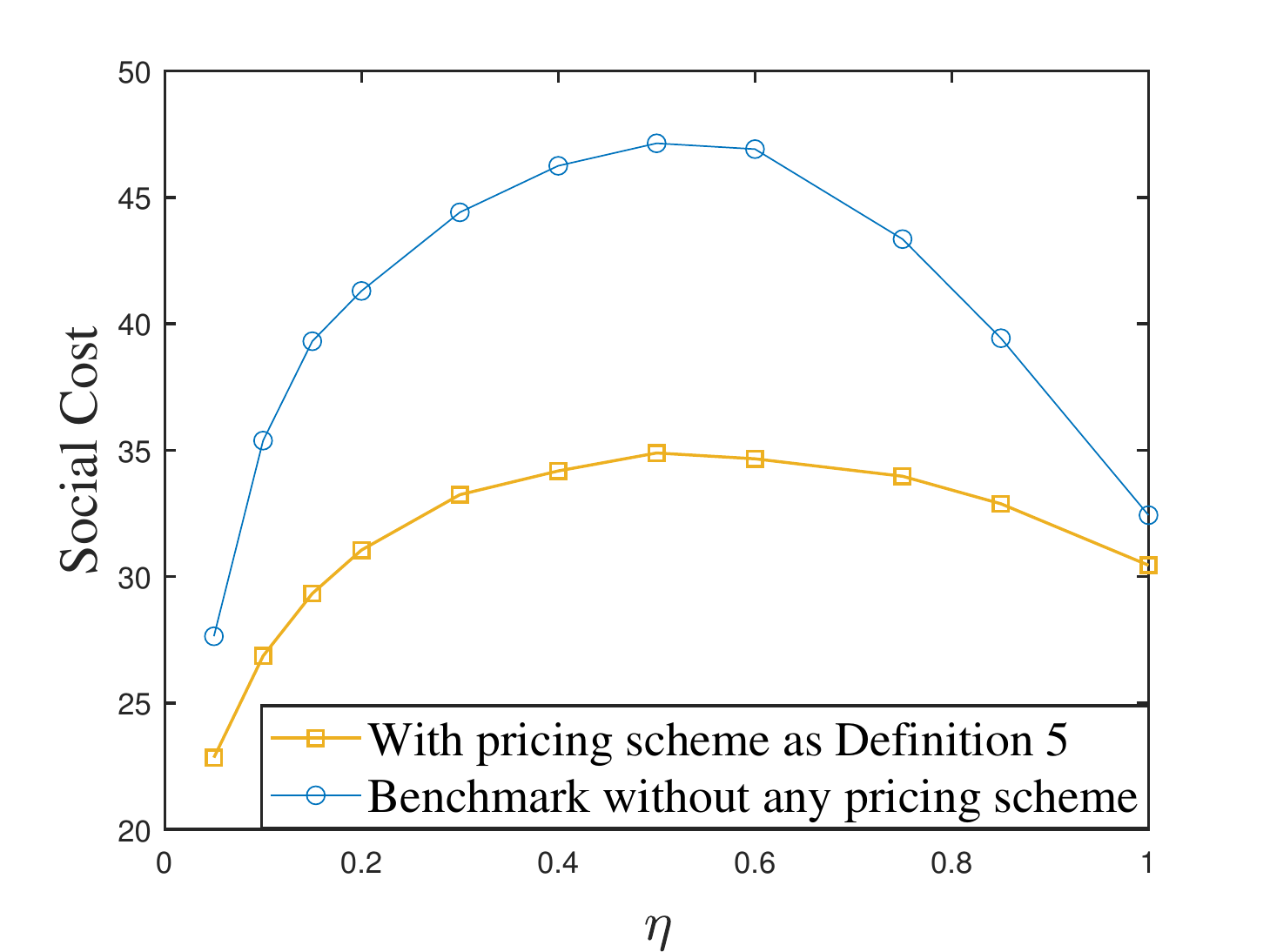}
\caption{Social costs with or without the pricing scheme 
in Definition \ref{Def:mechanism II} versus $\eta$.
The 
setting is $N=100$, $T=30$, $\alpha_L=0.25$, $\alpha_H=0.75$.
}
\label{Fig:social cost-II}
\end{figure}

When we fix 
$\alpha_L$ and $\alpha_H$, the parameter $\eta$ can be used to measure the divergence of clients' types $\alpha_i$'s. When $\eta=0.5$, the diversity of clients' type distribution is at its maximum. 
In our experiments on FL systems, we use a squared-SVM training model based on the original MNIST  dataset 
with a loss function $F_i\left(\mathbf{w}, \mathbf{x}_j, y_j\right)=\frac{\lambda}{2}\|\mathbf{w}\|^2+\frac{1}{2} \max \left\{0 ; 1-y_j \mathbf{w}^{\mathrm{T}} \mathbf{x}_j\right\}^2$ and $\lambda$ is a constant. The training output is a binary model, and each client has 1000 training and 1000 testing data samples in each simulation round.  
Fig. \ref{Fig:social cost-II} shows the social costs 
under our proposed pricing scheme in Definition \ref{Def:mechanism II} and the no-pricing benchmark. 
Both social costs increase as the variance of $\alpha_i$'s 
increases (i.e., $\eta \rightarrow 0.5$), and our pricing scheme more significantly reduces the social cost compared to the benchmark with the maximum , which are consistent with our findings under complete information.

\section{Conclusion}
In this paper, we adopt a game-theoretic perspective to examine clients' privacy-preserving strategies in federated learning systems. 
We propose a two-stage game model to capture the interactions between clients and the server, and introduce pricing mechanisms to regulate clients' selfish behaviors. Our pricing scheme is designed to reach the social optimum without verifying the actual noise-adding level.
Even under incomplete information of clients' private types, our mechanism incentivizes truthful reporting.
Our analysis of the ratio of social costs shows that our approach outperforms the unguided one, with simulation results revealing greater improvement as the diversity of clients' types increases.





\bibliographystyle{ieeetran}
\bibliography{ICCref}

\end{document}